\documentstyle[emulateapj,pstricks]{article}

\def\kms{\mbox{kms$^{-1}$}}
\def\kpch{\mbox{h$^{-1}$kpc}}
\def\mpch{\mbox{h$^{-1}$Mpc}}
     
\def\msunh{\mbox{h$^{-1}$M$_\odot$}}
\def\mv{\mbox{M$_{v}$}}
\def\mh{\mbox{M$_{h}$}}
\def\ome{\mbox{$\Omega_m$}}

\def\omel{\mbox{$\Omega_\Lambda$}}

\def\rv{\mbox{R$_{v}$}}

\def\vmax{\mbox{v$_{\rm max}$}}
\def\lnp{\mbox{LN$_p$}}
\def\lnn{\mbox{LN$_n$}}
\def\gp{\mbox{G$_p$}}
\def\gn{\mbox{G$_n$}}
\def\nbody{\mbox{$N-$body}}
\def\LCDM{\mbox{$\Lambda$CDM}}

\def\clone{\mbox{Cl$_{\rm Gp}$}}
\def\cltwo{\mbox{Cl$_{\rm LNp}$}}
\def\clthree{\mbox{Cl$_{\rm Gn}$}}
\def\clfour{\mbox{Cl$_{\rm LNn}$}}
\def\gone{\mbox{G1$_{\rm Gp}$}}
\def\gtwo{\mbox{G2$_{\rm Gp}$}}
\def\gthree{\mbox{G1$_{\rm Gn}$}}
\def\gfour{\mbox{G2$_{\rm Gn}$}}
\def\gfive{\mbox{G1$_{\rm LNp}$}}
\def\gsix{\mbox{G2$_{\rm LNp}$}}
\def\gseven{\mbox{G1$_{\rm LNn}$}}
\def\geight{\mbox{G2$_{\rm LNn}$}}

\def\mathnew{\mathsurround=0pt}
\def\ref{\par\noindent\hangindent=2pc \hangafter=1 }
\def\simov#1#2{\lower .5pt\vbox{\baselineskip0pt
    \lineskip-.5pt\ialign{$\mathnew#1\hfil##\hfil$\crcr#2\crcr\sim\crcr}}}  
\def\simgreat{\mathrel{\mathpalette\simov >}}

\def\'#1{\ifx#1i{\accent"13\i}\else{\accent"13#1}\fi}
\def\eg{e.g.,}

\begin{document}
\slugcomment{{\em submitted to the Astrophysical Journal}}

\lefthead{STRUCTURE OF HALOS IN NON-GAUSSIAN COSMOLOGICAL MODELS}
\righthead{AVILA-REESE ET AL.}

\title{The effects of Non-Gaussian initial conditions on the structure and 
substructure of Cold Dark Matter halos}

\author{Vladimir Avila-Reese and Pedro Col\'in}
\affil{Instituto de Astronom\'{\i}a, U.N.A.M., A.P. 70-264, 04510, M\'exico, 
D.F., M\'exico}

\author{Gabriella Piccinelli}
\affil{Centro Tecnol\'ogico, ENEP Arag\'on, UNAM. Av. Rancho Seco s/n, Col. 
Impulsora, Cd. Nezahualc\'oyotl, M\'exico}

\author{and}

\author{Claudio Firmani}
\affil{Osservatorio Astronomico di Brera, via E.Bianchi 46, I-23807 Merate, Italy}

\keywords{cosmology:dark matter --- galaxies:formation --- galaxies:halos --- 
methods:\nbody\ simulations}

\begin{abstract}

We study the structure and substructure of halos obtained in \nbody\ 
simulations for a $\Lambda$ Cold Dark Matter (\LCDM) cosmology with 
non-Gaussian initial conditions (NGICs). The initial statistics are lognormal
in the gravitational potential field with positive (\lnp) and negative (\lnn)
skewness; the sign of the skewness is conserved by the density field,
and the power spectrum is the same for all the simulations. 
Our aim is not to test a given non-Gaussian statistics, but
to explore the generic effect of positive- and negative-skew statistics
on halo properties.  From our low-resolution simulations, we find that
\lnp\ (\lnn) halos are systematically more (less) concentrated than
their Gaussian counterparts. This result is confirmed by
our Milky Way- and cluster-sized halos resimulated with high-resolution.
In addition, they show inner density profiles that depend on the 
statistics: the innermost slopes of \lnp\ (\lnn) halos are steeper 
(shallower) than those obtained from the corresponding Gaussian halos. 
A subhalo population embedded in \lnp\ halos 
is more susceptible to destruction than its counterpart inside Gaussian halos.
On the other hand, subhalos in \lnn\ halos tend to survive longer
than subhalos in Gaussian halos. The spin parameter probability distribution
of \lnp\ (\lnn) halos is skewed to smaller (larger) values
with respect to the Gaussian case. Our results show how the statistics of the 
primordial density field can influence some halo properties,
opening this the possibility to constrain, although indirectly, the primordial
statistics at small scale.

\end{abstract}
 

\section{Introduction}


The inflationary-motivated Cold Dark Matter (CDM) cosmology
describes rather well a large body of observational data at
large scales and has served as an useful theoretical framework
for investigating galaxy formation and evolution (see for 
recent reviews e.g., \cite{F02}; \cite{FA03}). 
In the CDM cosmology, the initial density fluctuation field is 
the starting point for calculating the structure formation in the 
universe. The simplest assumption for 
the statistical distribution of the primordial density fluctuations 
is the Gaussian one, which is valid only in the limit of zero
fluctuation amplitude; for finite rms fluctuations, the Gaussian
statistics assigns a non-zero probability to regions of negative
density. In the context of simple inflationary models, the Gaussian 
distribution may arise in a natural fashion. However, as we will 
see below, deviations from Gaussianity are possible, in particular 
for more complicated models of inflation, while from the observational 
point of view the results regarding non-Gaussianity are controversial. 
Moreover, it may happen that the statistics depend on scale, so that 
measurements made at large scales are not necessarily representative 
of the small scale statistics.
An interesting question to explore is how non-Gaussian initial 
conditions (NGICs) do affect cosmic structure formation at small scales.  
In this paper, we are interested in particular in the effects of 
NGICs on the density profiles of galaxy- and cluster-sized CDM halos 
as well as on the substructure abundance in the halos.

Following, we present a short review of theoretical studies
and observational constraints related to non-Gaussianity, as well 
as a brief discussion of previous numerical works on large-scale 
structure formation using  NGICs.
 
\subsection{Generation of Non-Gaussian fluctuations}

In the context of inflationary models, non-Gaussian fluctuations would 
be the signature of non-linearities during inflation. For the simplest 
single scalar field models, the slow-roll conditions prevent non-Gaussian 
primordial fluctuations from developing significantly (\cite{KL87}; 
\cite{OLM88}; \cite{MOL89}; \cite{BC90}; but see \cite{ABMR02}, who suggest that
non-Gaussianity may develop at second order calculation in cosmological 
perturbations). 
These constraints can be partially relaxed when special
non-linear conditions such as a non-vacuum initial state for the inflaton
(\cite{MRS00}; \cite{GMS02}) or a departure from scale invariance
in the inflaton potential (\cite{LPS98}; \cite{S92}) are introduced.
Deviations from a scale invariant 
spectrum indeed can make the primordial non-Gaussianity non-negligible 
(Acquaviva et al.). There are some pieces of evidence that the measured matter 
power spectrum presents a peculiar 
behavior: a sharp peak or break at $k \approx 0.05 \mpch$ 
(e.g., \cite{E97}; \cite{P97}; \cite{GB98}), an 
abrupt bending of the spectral index $n_s$ at scales smaller than 
$k \approx 0.1 \mpch$ (\cite{P97}), and/or a runing $n_s$ from $n_s > 1$ at 
large scales to $n_s < 1$ at small scales (\cite{P03}).

However, significant deviations from Gaussianity are more probable to occur 
in multi-field inflationary models. In this case, the deviations are produced
by fluctuations of the field that does not drive inflation, and 
it is consequently not constrained by a flat potential. Generically, one 
gets a mixture of adiabatic and isocurvature perturbations, quadratic in a 
Gaussian field (e.g., \cite{AGW87}; \cite{P99}; 
\cite{AMM97}; \cite{LM97}).
In some proposals, the mixing between adiabatic and isocurvature modes may 
effectively transfer the non-Gaussian features from one sector to the other,   
rendering them more significant (\cite{GWBM01}, and references 
therein; \cite{BU02}; \cite{BMR02}). 
An overview of the characteristics and limitations of the non-Gaussian 
contributions in the different inflationary models, either with a single or 
with several scalar fields is presented by Bernardeau \& Uzan.

A recently proposed alternative scenario consists in adiabatic 
density perturbations which originate after inflation from the perturbation 
of a field that is not the inflaton, and does not dominate the dynamics,
the so called {\it curvaton} (\cite{LW02}; \cite{LUW03}; see also 
\cite{LM97}; Moroi \& Takahashi 2001, 2002). The 
perturbation is initially of the isocurvature type but, if the curvaton is 
long-lived, its energy density will have a relative increase with respect to 
radiation, and will finally generate a curvature perturbation. The curvaton 
scenario can generate significant non-Gaussianity since the curvature 
perturbation is proportional to the perturbation in the curvaton energy 
density, which is in turn a linear combination of the curvaton-field 
perturbation and its square. 
In this model, large isocurvature perturbations, correlated with adiabatic 
perturbations, may arise in dark matter and baryonic matter. 

In contrast, in the topological-defect model, strong non-Gaussian
(skew-positive) fluctuations in the density field rather than in the
potential field may develop (e.g., \cite{TS90}; \cite{ASW98}).
Although this model has proved to be inconsistent with observations 
(e.g., \cite{PST97}; \cite{DKM98}; \cite{A97}), 
a mixed model of inflation plus certain defects (cosmic strings and any other
kind of global defects) seems to be still allowed by observations (\cite{B02}; 
\cite{Sk02} and references therein).

\subsection{Observational constraints}

As it has been just discussed, the theoretical prediction of the 
exact amplitude and precise 
form of non-Gaussianity in inflation is still an open question. Thus, we 
should turn to observations to look for its signatures and constraints (see 
\cite{Kamion03} for a recent review on the subject). Care must be taken in 
drawing conclusions about the Gaussian nature of the density field from 
observations; the uncertainty in the measurements and the limited coverage 
of present observations can hide cosmological non-Gaussian features. We should 
also be aware that the gravitational evolution and some systematic effects 
introduce non-Gaussian signatures in an intrinsically Gaussian distribution. 
We also would like to stress that there are no {\it a priori} strong 
arguments to postulate that the statistics of the primordial fluctuations 
should be the same at all scales.

The most direct way to search for primordial non-Gaussianity (at large scales)
is the Cosmic Microwave Background Radiation (CMBR). Many tests applied 
to the CMBR data resulted nearly consistent with Gaussianity (see e.g., 
\cite{K96}, and \cite{BT00}, and references therein, for COBE 
DMR maps; \cite{P02} for BOOMERanG maps; \cite{S02}, and \cite{W01} 
for MAXIMA maps; \cite{P01} for QMAP and Saskatoon maps; 
\cite{K03} for WMAP results). However, non-Gaussian statistics have 
no generic signatures on the sky and different tests and data analysis, at 
different scales, may be better or worse suited for different types of 
non-Gaussian statistics. Therefore, it is important to consider carefully 
any new result that supports the possibility of non-Gaussian features using 
different techniques and indicators (e.g., \cite{FMG98}; 
\cite{M00}; \cite{PVF98}; \cite{CNVW03}). 
Notice that \cite{BZG00} have shown that the 
non-Gaussian signal detected in the COBE data by Ferreira et al. was 
probably due to a systematic period in the COBE satellite flight 
---the ``eclipse period''. See also \cite{B00} for some comments 
on the other mentioned works.

The difficulty of predicting the precise form of non-Gaussianity from inflation or 
other seed mechanisms, led to define and work with some parametric indicators 
of the deviation from Gaussianity in a generic model instead 
of trying to look for a particular statistics. An often employed indicator is 
the non-linear coupling parameter $f_{\rm NL}$ that parameterizes the leading 
order non-linear/non-Gaussian correction to the primordial gravitational potential: 
$\Phi = \phi_L + f_{\rm NL} [\phi_L^2 - <\phi_L^2>]$, where $\phi_L$ is a Gaussian 
random field. A similar definition is introduced for non-Gaussianity in 
the density field (see e.g., \cite{V00}). The recent first-year WMAP 
observations (processed with two statistics: the angular bispectrum and 
Minkowski functionals) have constrained $f_{\rm NL}$ in the range 
$-58 < f_{\rm NL} < 134 \ \ (95\%)$ (\cite{K03}). As a reference, let 
us recall that in some non-standard inflationary models,
as the multi-field ones, $f_{\rm NL}$ can be as big as $\sim 20$ (Verde 2001), 
while in the single-field models with some non-linearities, 
$|f_{\rm NL}|\lesssim 1$ (\cite{ABMR02}). 
Combining WMAP data with ACBAR, 2dF, and HST results, strong constraints 
can be obtained in specific cases of correlated primordial and adiabatic 
density fluctuations, such as  curvaton models (\cite{GL02}).

\cite{TOH03} have performed an independent
foreground analysis from the WMAP data. Their foreground-cleaned and 
Wiener-filtered maps were used by Chiang et al. (2003) for a non-Gaussianity 
test. These authors implement a phase-mapping technique (Chiang, Coles \& 
Naselsky 2002), which has the advantage of testing non-Gaussianity at separate 
multipole bands (angular scales). A multipole band which is considerable
non-Gaussian could have an insignificant non-Gaussian contribution to the whole
map and produce thus an overall Gaussian realization. The authors show
that the foreground-cleaned map is against Gaussianity at all 4 multipole
bands centered around $\ell = 150, 290, 400,$ and 500. The Wiener-filtered
map, on the other hand, is Gaussian for $\ell < 250$ ($\simgreat 180$\mpch)
but non-Gaussian for $\ell > 350$ ($\lesssim 50$\mpch). As it was said above,
the statistics might change with the scale.  

The effects of primordial non-Gaussianity can be also studied in the
distribution of mass in the Universe (intermediate scales). The gravitational 
evolution not only amplifies the primordial density fluctuation
field but also makes it non-Gaussian due to clustering. However, the 
non-Gaussianity generated by gravitational evolution can be calculated and 
subtracted from the present statistics (e.g., \cite{V00}). Some of the tests of 
non-Gaussianity include the abundance and distribution of galaxy clusters 
and galaxies, the evolution of these abundances, and the size-temperature 
relation of galaxy clusters.

Since galaxy clusters form at the highest  density peaks, which are most 
sensitive to deviations from Gaussianity, their abundance, and in particular 
the evolution of their abundance, can be used to probe the statistics. A 
density distribution with positive (negative) skewness leads to more (less) 
high density peaks. In this field, a useful indicator of the amount of 
deviation from Gaussianity is the probability of obtaining a peak of 
height $3 \sigma$ or higher (at a given ``filtered'' scale) for a certain 
statistics relative to the Gaussian one: 
\begin{equation} \label{g3}
G_3 = 2 \pi {\int_3^{\infty} P(y) dy \over \int_3^\infty e^{-y^2 / 2} dy}.
\end{equation}
\cite{RGS00}, and \cite{KST99} (see also \cite{COS98}; 
\cite{W00}) constrain primordial 
non-Gaussianity through the evolution of cluster abundance. They use a 
Press-Schechter formalism extended to NGICs for predicting cluster abundances
at different redshifts and for different cosmological models.
Koyama et al., using data sets at $z = 0$ and $z \sim 0.6$ for cluster 
abundance and the correlation length from the APM survey, obtained the upper 
bound $\Omega_m < 0.5$ and the lower bounds $G_3>2$ and $\sigma_8 > 0.7$ for 
a $\chi^2$ non-Gaussian distribution; in particular for $\Omega_m = 0.3$ and 
$\Omega_{\Lambda} = 0.7$, non-Gaussianity of the order of $G_3\sim 4$ (skew 
positive) is favored. Robinson et al. (2000), from several samples of clusters 
at redshifts between 0.05 and 0.8, find that for $\Omega_m = 0.3$, Gaussianity 
is consistent with the data, but a wide range of non-Gaussian lognormal models 
also fit the data, with values of $G_3 < 4$ ($G_3 < 6$ when including cosmological 
constant) acceptable at the 2$\sigma$ level. It must be stressed that current 
constraints are not conclusive due to the difficulty in the determination of 
the mass of high-redshift clusters.

A long non-Gaussian tail would give not only an enhanced cluster abundance, 
but also a larger scatter in the formation redshifts and sizes of clusters 
with a given mass. A skew-positive distribution implies that halos 
collapse during a broader range of redshifts than in the Gaussian case, 
leading to a larger scatter of clusters properties, such as size and temperature. 
\cite{V01a} find that the predicted scatter for Gaussian initial 
conditions is consistent with the observed one and constrain $G_3$ to 
be $\lesssim 4$. 

Galaxies at high redshift are another kind of rare objects in the universe, 
so their abundance is considered for testing non-Gaussianity by \cite{V01b}. 
Results are still very preliminar and uncertain.
On the other hand, the bispectrum of the galaxy distribution 
from 2dF (\cite{V02}) and from  SDSS (\cite{Sz02}) surveys 
have been found consistent with Gaussian initial conditions.

Finally, the statistics of large-scale galaxy clustering can also be used to 
constrain non-Gaussianity. Most of the previous cosmological numerical simulations 
with NGICs were aimed at studying precisely galaxy clustering. In the next subsection
we review some of these works. In the present paper, we will introduce a 
new potential test for cosmological models with NGICs at the smallest scales, 
by exploring the inner structure and substructure of the dark matter halos.

\subsection{Previous numerical works}

In an attempt to overcome the problem of lack of power at large-scales of 
the standard CDM cosmology, among other alternatives, models with NGICs were 
proposed and tested by means of N-body simulations.
Moscardini et al. (1991, hereafter MMLM; see also 
\cite{MMLM90}) constructed certain multiplicative non-Gaussian statistics 
for the peculiar gravitational potential, and used them as initial 
conditions of \nbody\ cosmological simulations. The outcome of the 
simulations were analyzed by different statistical tests at large-scales 
and compared with observations. It was found that both the clustering dynamics 
and the present day texture are mostly sensitive to the sign of the 
initial skewness of density fluctuations. Positive models cluster to 
a lumpy structure with small coherence length, while negative models 
build up a cellular structure with large coherence
length, large voids, and enhanced peculiar motions. In a subsequent paper,
the same authors showed that some relevant differences may appear for models
with the same sign of the skewness but different statistics (\cite{CMPLMM93}).
For example, they showed that their CDM skew-negative $\chi^2$ model agrees
rather well with the 2D topology inferred from the Lick Catalogue, while the
skew-negative lognormal model (as well as the skew-positive models)
give discrepant results. 
From the analysis at smaller scales, \cite{MLMM91} found that 
groups in skew-negative models are preferentially loose systems, slightly biased 
over the background, with high velocity dispersion and clear signal 
alignment with neighboring groups. Groups in skew-positive models are highly 
biased compact systems with lower velocity dispersion.
In general, in the series of papers mentioned above, the authors
concluded that standard CDM skew-negative models (predominance of
primordial underdense regions) retain the advantages of the Gaussian 
biased standard CDM counterpart, and improve on several of their problems.

An extensive numerical study of the influence of NGICs on large-scale
structure formation has been done also by \cite{WC92}. 
These authors attempted to isolate the effects of NGICs from other
factors by considering a range of power-law power spectra and different 
cosmological models. Similar to \cite{MLMM91}, they found that 
skew-positive models form structure by accreting mass onto rare high peaks, 
leaving large volumes almost unperturbed, while skew-negative models develop
structures rich in expanding voids, sheets and filaments. The authors
conclude that skew-positive models produce unrealistic large-scale
features, while skew-negative models create attractive structure
but, for the low-density universe ($\Omega_m = 0.2$) cluster velocity
dispersions are too low and the voids are probably excesively large.

It should be noted that since the time these works appeared (early 90's), 
deep progresses in the determination of the cosmological
parameters, the observational data on galaxy clustering, and 
the computational capability were made. Time is ripe to explore the 
effects of NGICs on stucture formation by means of high-resolution 
\nbody\ simulations for the preferred (concordance) cosmological model.  
This paper presents by first time a numerical study about the structure of 
halos formed under NGICs.

In \S 2 we present the non-Gaussian models to be used here, while in \S 3
the implementation of the NGICs in the \nbody\ code is described
and the  different low- and high-resolution runs are presented. In 
\S 4.1, the halo concentrations and density profiles for 
the non-Gaussian and Gaussian runs are presented and compared among them. The 
spin parameter distribution of the low-resolution halos and the 
angular momentum distribution of the high-resolution halos is 
analyzed in \S 4.2. The ellipticity of the latter halos is analyzed
in \S 4.3. In \S 4.4, results on the subhalo population in the 
galaxy and cluster halos are presented. The summary and the discussion are 
given in \S 5.

\section{Non-Gaussian models}

Since there is no preference for any particular non-Gaussian statistics
(\S 1.1), we chose arbitrary non-Gaussian models. Our main interest
is to study generically the influence of skew-positive and -negative NGICs 
on halo properties. To get the initial conditions for our simulations, we 
have followed closely MMLM, who constructed the non-Gaussianity in the 
gravitational potential fluctuation field, $\Phi$, instead of the density
field. The potential fluctuation $\Phi$ is related to the linear mass 
fluctuation via the Poisson equation. We first generate a Gaussian 
realization in Fourier space of a $w$ random field with the primordial 
(scale-invariant) power spectrum
\begin{equation}
P_w (k) = B k^{-3},
\end{equation}
where $k$ is the comoving wave number, and $B$ is the normalization factor. 
The zero-mean Gaussian field $w$ is obtained in configuration space by 
inverse-Fourier transforming $\tilde{w}(\mbox{\boldmath $k$})$. The 
non-Gaussian field for the perturbations in the potential, 
$\Phi(\mbox{\boldmath $x$})$, is then built as the convolution of a real 
function $\tau(\mbox{\boldmath $x$})$ with a stationary, zero-mean random
field $\phi(\mbox{\boldmath $x$})$, which is obtained from a non-linear 
transformation of the $w(\mbox{\boldmath $x$})$ Gaussian field. 
For a lognormal statistics the relation between $w$ and the 
primordial potential fluctuation $\phi$ is given by
\begin{equation} \label{fluct}
\phi(\mbox{\boldmath $x$}) = A \left[ e^{w(\mbox{\boldmath $x$})} - \left< e^w \right> \right],
\end{equation}
where $A$ is a normalization factor, related to $B$ in eq. (2). The Fourier 
transformed of the function $\tau(\mbox{\boldmath $x$})$ is the transfer 
function $T(k)$ used to get the processed power spectrum at the recombination 
epoch. The Fourier transformed function $\tilde{\Phi}(k)$, which is simply given
by the product of $\tilde{\phi} (k)$ with  $T^2(k)$, is subsequently 
processed by the Zel'dovich approximation algorithm.

Different lognormal statistics are obtained depending on
the amplitude of the $w$ Gaussian field. Although MMLM 
set $\left< w^2 \right> = 1$, we prefer
to assume a variance slightly smaller than one 
because it reproduces better the theoretical \LCDM-density 
power spectrum. There is still freedom in
choosing the sign of the constant $A$ in eq. (3),
with the sign determining the direction of the skewness. 
Our Gaussian realizations are also implemented using a 
$\phi$ field but in this case
\begin{equation}
\phi(\mbox{\boldmath $x$}) = A w(\mbox{\boldmath $x$}).
\end{equation}
We notice that because Gaussian models with oposite
signs give rise to different realizations in a statistical sense, 
we had to simulate for each lognormal skew-positive \lnp\ or lognormal 
skew-negative \lnn\ model the corresponding gaussian
\gp\ or \gn\ model. We also notice that because in general
a statistical distribution for the gravitational
potential does not imply the same distribution for the
density field (MMLM), we should not expect
the statistics of the density field to be lognormal,
nevertheless the sign of the skewness is preserved.

In order to compute the probability distribution
function (PDF) for the different statistics at 
the initial time $z_i
= 50$ (after the application of the Zel'dovich
approximation), our 100\mpch\ box simulation 
(see below) was divided in 64$^3$ cells. The 
density in each cell was then computed using the cloud-in-cell 
scheme. The skewness parameter 
$\xi \equiv \left< \delta^3 \right> / \left< \delta^2 \right>^{3/2}$ 
is 0.52, 3.90, and -1.27 for the Gaussian, \lnp\, and \lnn\
statistics, respectively, while the $G_3$ parameter defined in
eq. (1) is 3.05 for \lnp\ and 0.067 for \lnn. Note that
at $z_i$ the PDF of the density field of a Gaussian realization is not
Gaussian; in fact, we found that the PDF could be fitted by
a lognormal, skew-positive, distribution: non-linear
modes appear as a result of the application of the
Zel'dovich approximation. This is why
the skewness of the density field in the Gaussian 
scenario is not zero anymore. 

\section{Numerical simulations}

The Adaptive Refinement Tree (ART) \nbody\ code (\cite{KKK97}) 
in its multiple-mass version was used to run a series of
simulations aimed mainly at studying the structure of halos under different 
statistics.  The ART code achieves high
spatial resolution by refining the base uniform grid in all
high-density regions with an automated refinement algorithm.
The multiple-mass scheme is used to increase the mass and spatial 
resolution in few selected halos and its implementation is
described in detail elsewhere (\cite{Klypin01}).
All \nbody\ simulations were performed in a low-density flat $\Lambda$CDM
model with the following parameters: $\Omega_m=0.3$,
$\Omega_{\Lambda}=0.7$, and $h=0.7$. The power spectrum was
normalized to $\sigma_8 = 1.0$, where $\sigma_8$ is the $rms$ mass
fluctuation in spheres of radius $8 \mpch$ (the top-hat window is used). 

As a test of our method for generating NGIC's, we first ran two 
simulations under \gp\ and \lnp\ statistics
in a box of 20 \mpch\ for each side and 64$^3$ particles. Halos from these
simulations were also used for further analysis. We also ran
two simulations with low-mass resolution in boxes of 12.5 and 100
\mpch\ on a side. We selected, for resimulation with high-mass
resolution and for each statistics, two galaxy-sized halos from the
12.5\mpch\ box and one cluster-sized halo from the 100\mpch\ box. 
In an attempt to study galaxy-sized halos with different mass
assembly histories (MAHs), we simulated with high resolution two halos, 
$G1$ and $G2$, with different concentrations
since it is known that the concentration of the halo is related to its 
MAH (Avila-Reese et al. 1998, 1999; \cite{WBPKD02}). We notice, 
however, that in a box as small as the one we chose for the
search of galaxy-sized halos, only two or three halos can be found
to satisfy simultaneously the following criteria:
(a) comparable mass, (b) extreme concentrations,
(c) isolation, and (d) with a counterpart in the corresponding lognormal 
realization. Halos \gone\ and \gtwo, for example, obey roughly these 
rules. They have masses that differ by a factor of three and NFW (
\cite{NFW97}; \cite{ENS01}) concentrations, 
$c_{\rm vir}$, of 16.4 and 11.4, for \gone\ and
\gtwo, respectively. The reason why we did not find a galaxy-sized halo
with a very low concentration is because very often these halos
are found accompanied with halos of comparable mass. In the rare case
in which we do find an isolated halo with a low $c_{\rm vir}$, it turns out
to be too small or to lack a ``partner'' in the lognormal statistics.
\begin{figure*}[ht]
\pspicture(0.5,-1.5)(13.0,16.0)
\rput[tl]{0}(2.8,15.0){\epsfxsize=15cm
\epsffile{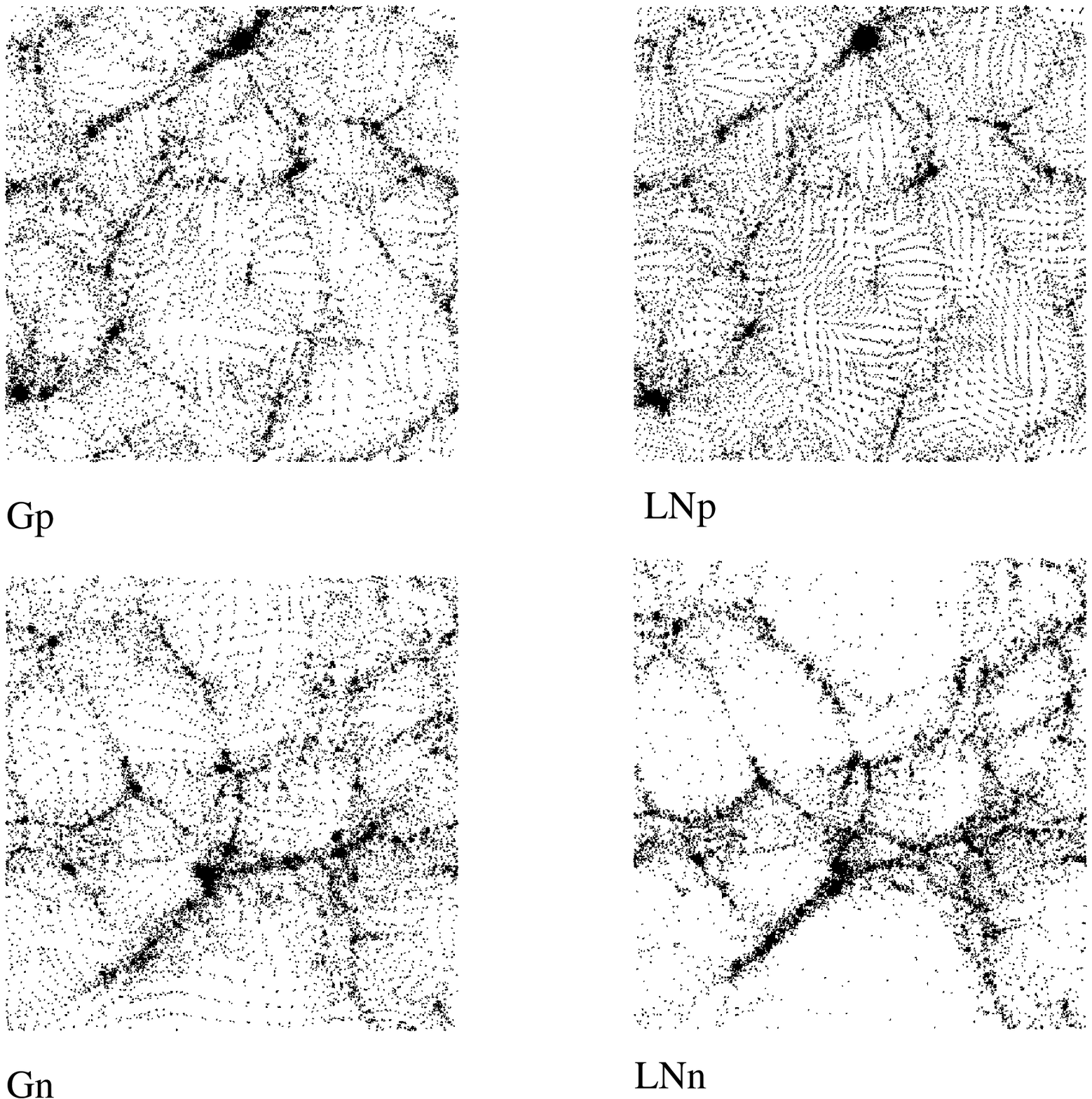}}
\rput[tl]{0}(0.5,1.0){
\begin{minipage}{18.4cm}
  \small\parindent=3.5mm {\sc Fig.}~1.---
Projected particle positions in slices of one tenth the 
size of the 100 \mpch\ box at $z = 0$. Each slice contain about 26,000
dark matter particles. We note that the particle distributions are
compared {\it at the same cosmological time} and not when the clustering properties
of the different scenarios are comparable, as has often been the case 
in the literature (\eg\ MMLM). Although, the differences in the large-scale
pattern between a lognormal realization and its Gaussian counterpart are not
as striking as if they were compared {\it at the same clustering epoch}, we
do see some differences: voids are emptier and more defined in the \lnn\
than in their Gaussian scenario. On the contrary, in the \lnp\ scenario
voids are not as developed (they contain more DM particles) as in the
Gaussian one.
\end{minipage}}
\endpspicture
\end{figure*}

The bound density maxima (BDM) group finding algorithm (\cite{KH97}) 
was used to locate halos and subhalos in all simulations. 
The BDM algorithm identifies positions of local maxima in the
density field at the scale of interest and applies physically
motivated criteria to find out whether a group of particles forms
a gravitationally bound halo. 
\begin{planotable}{rrcccrcrc}
\tablecolumns{8}
\tablewidth{0pc}
\tablecaption{Simulation Parameters}
\tablehead{\colhead{statistics} & \colhead{$L_{BOX}$} & \colhead{$m_p$} & \colhead{Resolution}
& \colhead{$M_{vir}$} & \colhead{\vmax} & \colhead{c$_{1/5}$} & \colhead{$\lambda'$} & \colhead{Halo name tag} \\
& (\mpch) &  (\msunh) & (\kpch) & (\msunh) & (\kms) &          &($10^{-2}$) & }
\startdata
 \gp\ & 100.0 & $5.0 \times 10^9$ & 3.0 & $6.2 \times 10^{14}$ & 1407 &  6.1 & 2.82 & \clone\ \\
\lnp\ & 100.0 & $5.0 \times 10^9$ & 1.5 & $8.3 \times 10^{14}$ & 1643 &  7.2 & 1.37 & \cltwo\ \\
 \gn\ & 100.0 & $5.0 \times 10^9$ & 3.0 & $1.6 \times 10^{14}$ &  913 &  6.8 & 3.79 & \clthree\ \\
\lnn\ & 100.0 & $5.0 \times 10^9$ & 6.1 & $1.3 \times 10^{14}$ &  781 &  5.2 & 1.22 & \clfour\ \\
 \gp\ &  12.5 & $9.7 \times 10^6$ & 0.2 & $6.9 \times 10^{11}$ &  188 & 11.1 & 1.52 & \gone\ \\
 \gp\ &  12.5 & $9.7 \times 10^6$ & 0.2 & $2.3 \times 10^{12}$ &  241 &  8.0 & 1.22 & \gtwo\ \\
 \gn\ &  12.5 & $9.7 \times 10^6$ & 0.2 & $2.3 \times 10^{12}$ &  175 & 9.2  & 5.32 & \gthree\ \\
 \gn\ &  12.5 & $9.7 \times 10^6$ & 0.2 & $2.8 \times 10^{12}$ &  253 & 7.8  & 3.13 & \gfour\ \\
\lnp\ &  12.5 & $9.7 \times 10^6$ & 0.1 & $3.0 \times 10^{11}$ &  303 & 19.5 & 2.03 & \gfive\ \\
\lnp\ &  12.5 & $9.7 \times 10^6$ & 0.2 & $2.1 \times 10^{12}$ &  256 & 9.3  & 0.69 & \gsix\ \\
\lnn\ &  12.5 & $9.7 \times 10^6$ & 0.4 & $7.3 \times 10^{11}$ &  157 & 7.0  & 6.15 & \gseven\ \\
\lnn\ &  12.5 & $9.7 \times 10^6$ & 0.4 & $2.7 \times 10^{12}$ &  218 & 5.1  & 10.00 &  \geight\ \\
\enddata
\end{planotable}

An overview of some of the characteristics of the studied
halos are shown in Table 1. The symbol with
which we denote the statistics is shown in column (1). The 
mass of the least massive particle is shown in column (3). It is
only a function of the size of the box (col. [2]) because the
cosmological model and the number of mass species are fixed.
Column (4) shows the formal force resolution,
measured by the size of a cell in the finest mesh. In general,
the lognormal skew-positive (skew-negative) simulations have a
higher (lower) resolution than the Gaussian counterparts because
simulated halos in the former case reach a higher (lower)
central density than in the latter one. The mass \mv\ within
the virial radius, defined as the radius at which the average halo density 
is $\Delta_c$ times the background density according to the spherical top-hat
model, is shown in column (5). $\Delta_c$ is a number that depends on epoch and
cosmological parameters (\ome,\omel); for a flat $\Lambda$CDM model,
$\Delta_c \sim 337$ at $z = 0$, where $z$ is the redshift.
The maximum circular velocity defined as
\begin{equation}
        v_{\rm max} =\left( \frac{GM(<r)}{r} \right)^{1/2}_{\rm max},
\end{equation}
where $G$ is the gravitational constant and $M(<r)$ is the mass contained
within the radius $r$, is placed in column (6). In column (7) we show
the c$_{1/5}$ halo concentration parameter (see \S 4.1). The spin parameter 
$\lambda'$ as defined by \cite{Bullock2001} is shown in column (8) (see \S 4.2).
In column (9) we introduce the halo name tag for each one of the 
simulated halos. Cluster- and galaxy-sized halos are named $Cl_x$ and $Gn_x$,
respectively, where $x$ denotes the statistics and $n$ the halo number.

\section{Results}

Figure 1 shows projected particle positions at $z = 0$ from the 
100\mpch\ box low-resolution simulations in slices of
one-tenth of the box size.  The differences in the particle
distribution between the lognormal realizations and the
Gaussian ones are clearly seen:
the density distribution in the \lnp\ statistics is lumpier
with concentrated isolated peaks and a smaller coherence length,
while in the \lnn\ statistics it has a cellular structure with
bigger voids and a larger coherence length.
Notice that the void population is different in both cases; for the
\lnp\ model it appears to be larger than in the Gaussian case, while 
the contrary happens in the \lnn\ model. The contrasts among
the different particle distributions are not as remarkable as
shown, for example, in MMLM because of the different normalizations:
we normalize our models to the same $\sigma_8$ while MMLM stop
their simulations when ``the particle two-point function is
best fitted by a power law $\xi (r) = (r/r_0)^{-\gamma}$, with
$\gamma = 1.8$, in a suitable interval''; accordingly,
their \lnp\ (\lnn) models are less (more) evolved. We
measured $\xi (r)$ at $z = 0$ for our different statistics
and found, as expected, similar correlation length $r_0$ (defined as 
the radius $r$ where $\xi (r) = 1$) values. However, for $r < r_0$ the 
correlation functions of the different scenarios start deviating from 
each other, with the greatest differences found at the smallest scales. 
As expected, the $\xi$ functions of the \lnp, \gp\ (or \gn) and \lnn\ 
statistics are on the top, middle, and bottom, of the $\xi$ vs. $r$ 
diagram, respectively. In any case, it is not the aim of this paper to
explore the large-scale structure produced in simulations with NGICs,
but to explore the effects of NGICs on the properties of CDM halo.

Following, we present a study of the structural properties of
galaxy- and cluster-sized halos from the non-Gaussian realizations
and their comparison with the corresponding halos of the
Gaussian simulations.

\subsection{Concentrations and Density Profiles of Halos}

In Figure 2 we show several general concentration parameters (vs. the 
halo mass \mh) measured for the halos in the low resolution 
simulations (12.5, 20 and 100 \mpch\ box size).  These plots 
give us a preliminar idea of the effects of NGICs on the halo
properties for a relatively large number of objects. 
{\pspicture(-0.6,-0.5)(12.0,20.0)
\rput[tl]{0}(-0.7,19.5){\epsfxsize=8.5cm
\epsffile{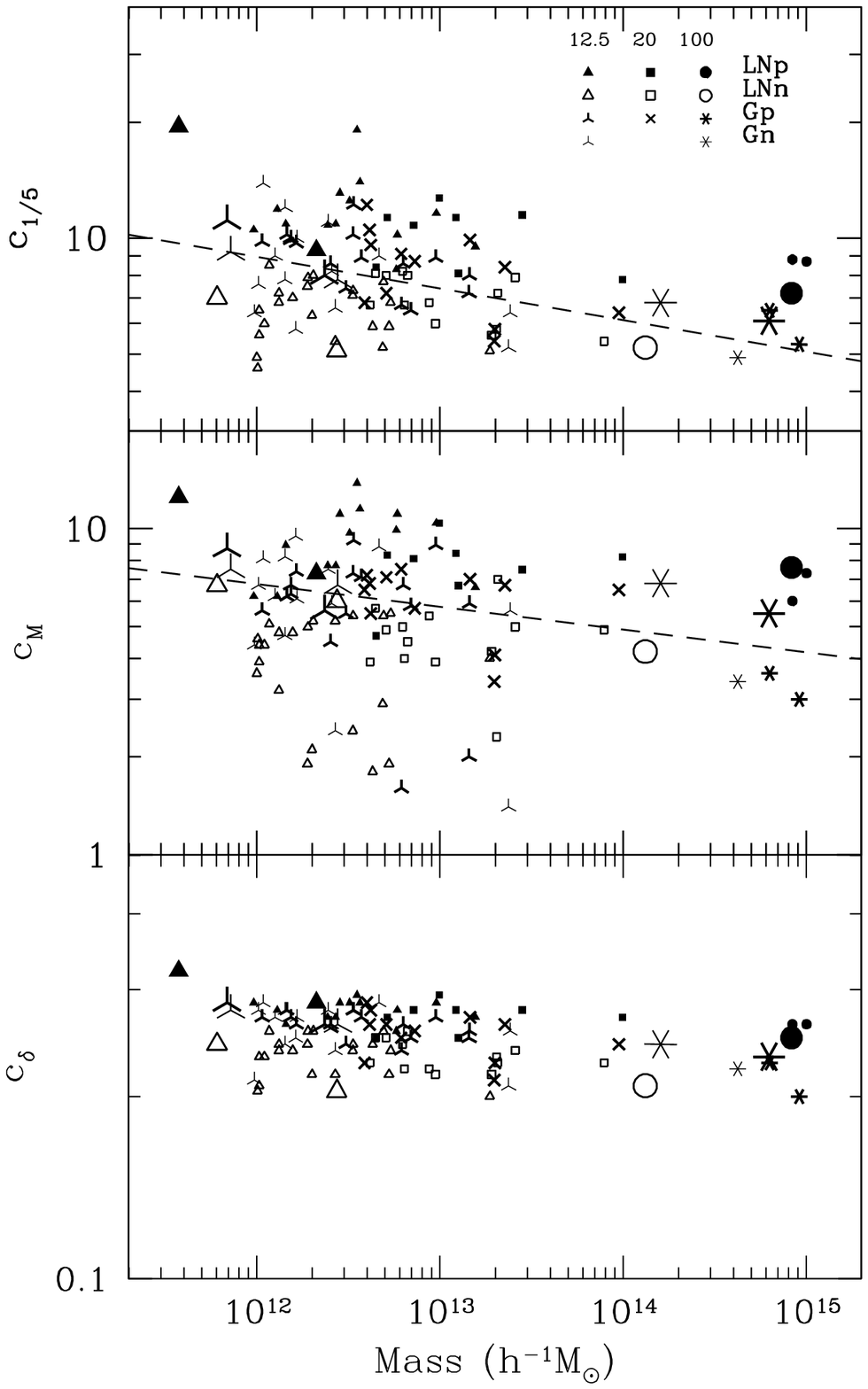}}
\rput[tl]{0}(-0.7,5.3){
\begin{minipage}{8.8cm}
  \small\parindent=3.5mm {\sc Fig.}~2.---
Different concentration parameters (see text) vs. \mv\ for 
the halos in the low-resolution simulations (small symbols) and those 
re-simulated with high-resolution (large symbols). Only halos with 
more than 1500 particles are plotted. The different symbols indicate
the box size and the statistics of the simulations (see the legends
in the upper panel). The dashed lines in the upper and medium panels are 
the linear fitting to the data (Gaussian isolated halos) presented in Avila-Reese 
et al. (1999), and the analytical estimate of c$_M$ assuming the NFW profile
and using log c$_{\rm  vir}= 2.35 - 0.1 log(\mv/\msunh)$. 
According to each one of the three concentration parameters used,
the \lnp\ (solid symbols) and \lnn\ (open symbols) halos are systematically
more concentrated and less concentrated than the Gaussian halos (squeletal
symbols), respectively.
\end{minipage}}
\endpspicture} 
Only halos with more than 1500 particles are plotted. The c$_{1/5}$ 
concentration
(upper panel) is defined as the ratio between the halo radius (virial or
truncation radius\footnote{If the halo density profile
flattens or even increases before attaining \rv, then the halo is truncated
at the flattening radius. This happens only in some cases for subhalos or
multiple halos. In most of the cases \rv is attained.}) and the radius 
where is contained one-fifth of the halo mass (\cite{AFKK99}). This parameter
depends on the definition and determination of the halo radius and mass. It 
is convenient to introduce other concentration parameters that do not depend on 
these definitions. Here we use the c$_M$ and c$_{\delta }$ parameters.
The c$_M$ concentration (medium panel) is defined as 27 times the ratio
between the mass at r$_{\rm in}=8.5$ kpc$\ (\vmax/220 \kms)$ and the mass
at r$_{\rm out}=3\times$r$_{\rm in}$ (\cite{Dave2001}).  
The c$_{\delta }$ concentration (lower panel) is defined as the ratio 
between the radius where the halo mean overdensity is $3\times 10^4$ and 
where it is $800$. The faster the density between these two radii decays
(larger concentration), the larger is c$_{\delta }$, tending to 1 in the 
limit. Notice that c$_M$ 
traces the halo concentration in the inner region, while c$_{\delta }$
refers to a large portion of the halo, it is more global than c$_M$.   
We do not use the more common NFW concentration parameter, c$_{\rm vir}$, 
because we do not know {\it a priori} whether the non-Gaussian halos will 
be described by the NFW density profile.

It is clear from Figure 2 that halos produced in the \lnp\ model are 
systematically more concentrated than the Gaussian ones, while the \lnn\ 
halos show an opposite trend. The \lnp\ halos from the different
box realizations are plotted with solid symbos, while those from 
the \lnn\ simulations are plotted with open symbols. The corresponding
Gaussian halos are plotted with skeletal symbols. Dashed lines are the 
expected values for NFW CDM halos. In the case of  c$_{1/5}$ we take the results
reported in Avila-Reese et al. (1999) for the isolated halos, while
for  c$_M$  we use its numerical relations with 
c$_{\rm vir}$, where for c$_{\rm vir}$ we took their average values in 
function of \mv\ as measured in the Gaussian $\Lambda$CDM simulation 
studied in Avila-Reese et al. (1999): 
logc$_{\rm vir}$\ =\ 2.35\ - \ 0.10\ log(\mv/\msunh)

Now we turn to the re-simulated halos with high-resolution. Their
concentrations are also shown in Fig. 2 with the corresponding large symbols. 
Figure 3 shows their density profiles. Upper and lower panels are for the 
skew-positive and skew-negative models, respectively. Left
and medium columns are for the high- and low-concentrated galaxy-sized
halos, respectively, and right column is for the single cluster-sized 
halo. The LN-model halo profiles are plotted with solid lines, while 
the corresponding Gaussian-model halo profiles (calculated with the
same seed) are plotted with dashed lines. The dotted lines
are the NFW fitting to the Gaussian halos; the fittings are satisfactory,
excepting at the very inner regions where the halo profiles are slightly 
steeper than the NFW profile, in particular for the halos in the lower panels.
As said in \S 2.2, we selected two different galaxy-sized halos from 
the low-resolution simulation in order to cover the intrinsical (cosmological)  
dispersion on halo structures, which is mainly due to the dispersion 
on the halo mass aggregation histories, given on its own by the statistical
nature of the primordial density fluctuation field.

\begin{figure*}[ht]
\pspicture(0.5,-1.5)(15.0,17.0)
\rput[tl]{0}(1.0,16.0){\epsfxsize=18cm
\epsffile{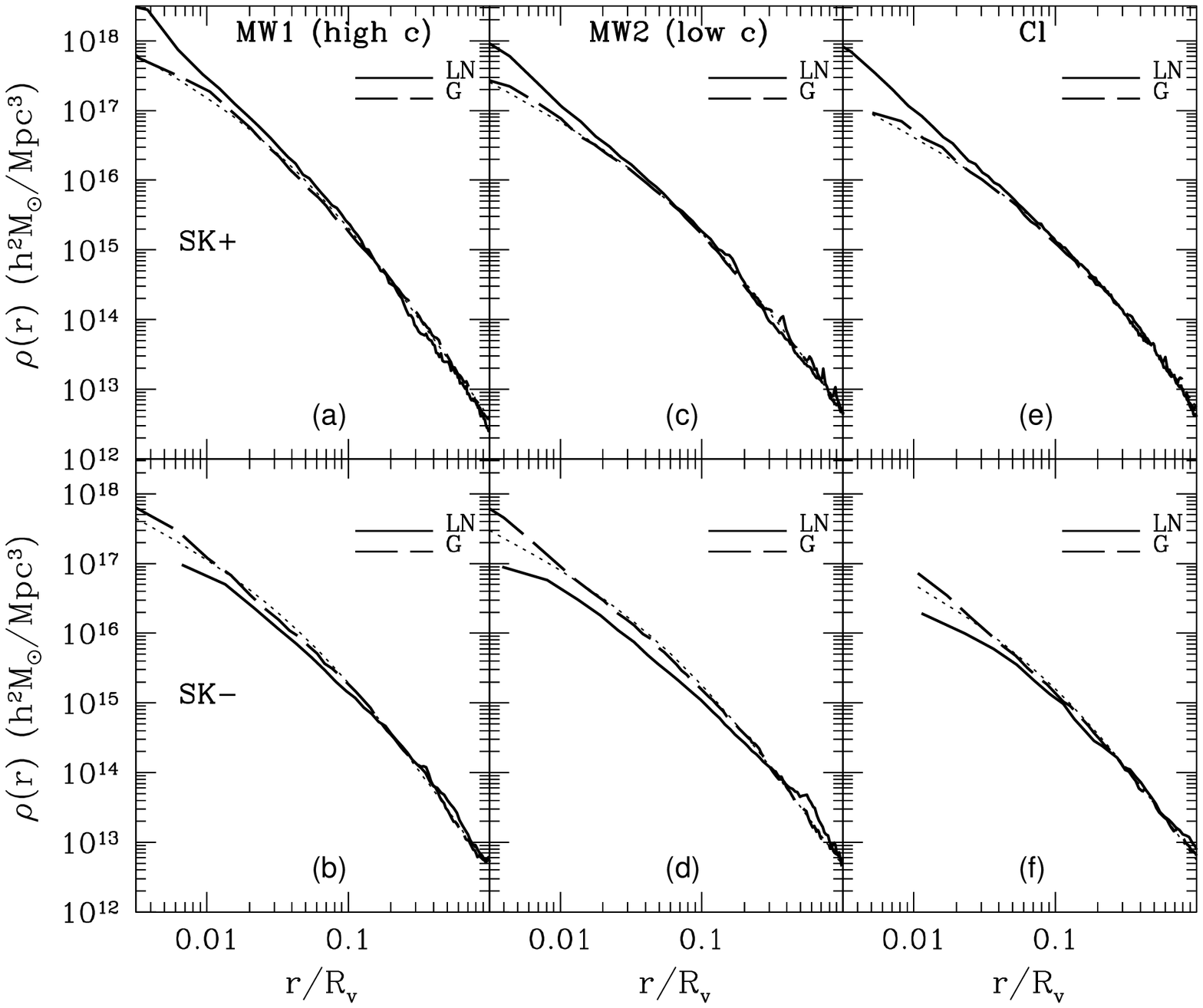}}
\rput[tl]{0}(0.5,0.0){
\begin{minipage}{18.4cm}
  \small\parindent=3.5mm {\sc Fig.}~3.---
Density profiles of halos re-simulated 
with high resolution. Upper and lower panels are for the skew-positive and
skew-negative statistics. The first and second columns are for the 
high- and low-concentration galaxy-sized halos, respectively, while the third
column is for the only simulated cluster-sized halo. Solid lines are for
the non-Gaussian halos and dashed lines are for the corresponding
Gaussian counterpart (simulated with the same random see). Dotted lines
are the best NFW fitting to the Gaussian-halo density profiles. The \lnp\
(\lnn) halos are more (less) cuspy than the Gaussian counterpart halos.
\end{minipage}}
\endpspicture
\end{figure*}

The results presented in Fig. 3 show that the NGICs have a clear
imprint on the density profile of the $\Lambda$CDM halos. For the 
\lnp\ model the halos are not only more concentrated than the 
corresponding Gaussian-model halos, but they are also significantly
cuspier. The opposite applies for the \lnn-model halos. In order
to see in more detail the features of the inner density profiles
of the halos, in Fig. 4 we show their slopes vs. the radius. Upper
and lower panels are for the skew-positive and skew-negative statistics.
In the left panels the two galaxy-sized halos are plotted (thick and 
thin lines are for the high and low concentration halos, respectively),
while the right panels are for the cluster halo. As in Fig. 3, solid
lines are for the non-Gaussian halos and dashed lines are for the 
corresponding Gaussian halos. In all the cases, the 
slopes of the non-Gaussian halos systematically depart
from the slopes of the corresponding Gaussian halos at 
radii smaller than $3-10\%$ the virial radius. The inner slopes
of the \lnp\ (\lnn) halos are smaller (larger) than those of their 
corresponding Gaussian halos. Notice that
the Gaussian halos corresponding to the \lnp\ run resulted
with shallower profiles than the Gaussian halos corresponding
to the \lnn\ run (this is part of the cosmic variance); the latter 
have inner slopes around -1.4. 
\begin{figure*}[ht]
\pspicture(0.5,-1.0)(15.0,16.0)
\rput[tl]{0}(1.5,15.0){\epsfxsize=16cm
\epsffile{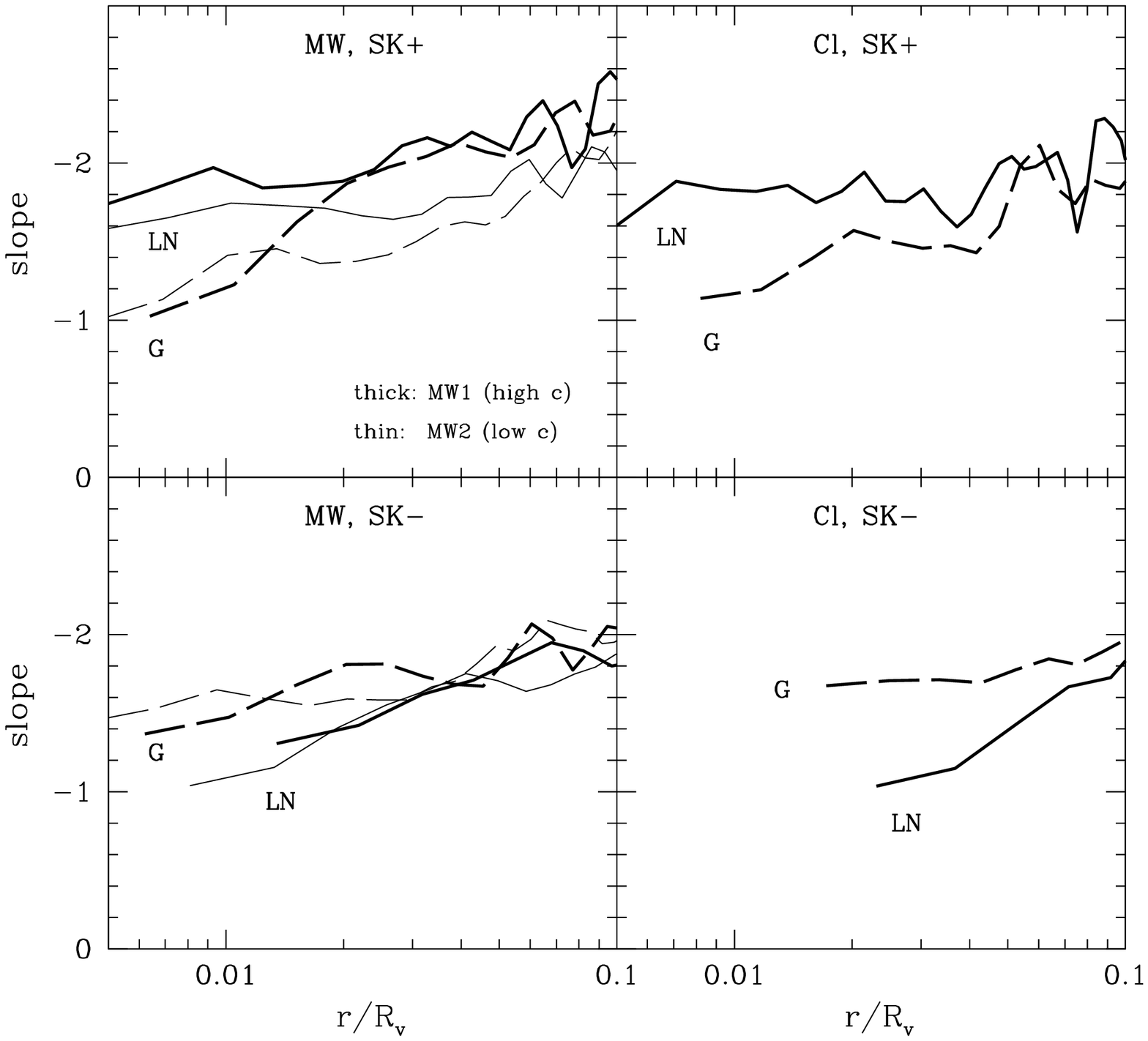}}
\rput[tl]{0}(0.5,0.0){
\begin{minipage}{18.4cm}
  \small\parindent=3.5mm {\sc Fig.}~4.---
Density profile slopes in the inner parts of the non-Gaussian
and Gaussian halos showed in Fig. 3. Upper and lower panels are for the 
skew-positive and skew-negative statistics. The two galaxy-sized halos are shown 
in the right panels (thick and thin lines for the high- and low-concentrated
halo, respectively), while the cluster-sized halo is in the right panels. The
line coding is the same than in Fig. 3.
\end{minipage}}
\endpspicture
\end{figure*}

\subsection{Spin Parameter and Angular Momentum Distribution}

In Figure 5 we show the probability distribution of the modified spin parameter, 
$p(\lambda')$,
for all halos from the 12.5 \mpch\ box with the different statistics; the
\gn\ and \gp\ halos were joined into one denoted by G (see top-left panel).
The parameter $\lambda'$ determines the global angular momentum of a 
halo and it is defined as in \cite{Bullock2001}:
\begin{equation}
\lambda' \equiv \frac{J_{v}}{\sqrt{2} M_{v} V_{v} R_{v}},
\end{equation}
where $J_{\rm v}$ is the angular momentum inside virial radius \rv,
and $V_{\rm v}$ is the circular velocity at radius \rv. In Table 1, column
(7), we show $\lambda'$ for the high-resolution halos. The 
halo angular momentum is calculated as:
\begin{equation}
\mbox{\boldmath $J$} = m_i \sum_{i=1}^{n} \mbox{\boldmath $r_i$}
\times \mbox{\boldmath $v_i$},
\end{equation}
where \mbox{\boldmath $r_i$} and \mbox{\boldmath $v_i$} are the position 
and velocity of the {\it i}th particle with respect to the halo
center of mass. 

The spin parameter distributions shown in Fig. 5 were obtained using halos 
with more than 100 particles. An extra simulation with $256^3$ particles, 
with the same parameters as the 12.5 \mpch\ box simulation described in 
\S 3, was also run and stopped
at $z = 1$. The about 1000 halos with more than 500 particles from this
simulation were used to compute a well resolved \lnp\ spin-parameter
distribution (top-right panel). A complete analysis of this simulation has
been deferred to a future paper. For Gaussian initial conditions, \nbody\ 
simulations have shown previously that $p(\lambda')$ can be well described by 
the lognormal distribution
(\eg\ Bullock et al. 2001),
\begin{equation}
p(\lambda')= \frac{1}{\sigma_{\lambda'} \sqrt{2 \pi}} 
\exp \left[-\frac{\ln^2 \left( \lambda'/\lambda'_0 \right)}
{2 \sigma_{\lambda'}}\right] \frac{d \lambda'}{\lambda'}.
\end{equation}
We confirm this for our Gaussian simulations and find that the halo spin 
parameter distribution in a lognormal statistics (with positive or negative 
skewness) is also roughly lognormal.
Curves on panels are the lognormal best-fits to the data: 
($\sigma_{\lambda'}, \lambda'_0$)=(0.63,0.035), (0.61,0.029), (0.73,0.026), 
and (0.49,0.042) for Gaussian, \lnp\ (high resolution), and \lnp\ and \lnn\
(low resolution) simulations, respectively. The figure shows that halos
formed in a \lnp\ statistics are biased to lower values of $\lambda'$; on 
the contrary, halos in the \lnn\ statistics seem to have on average a 
higher $\lambda'$ value than their Gaussian counterparts.
 \begin{figure*}[ht]
\pspicture(0.5,-1.0)(15.0,18.0)
\rput[tl]{0}(2.0,17.5){\epsfxsize=16cm
\epsffile{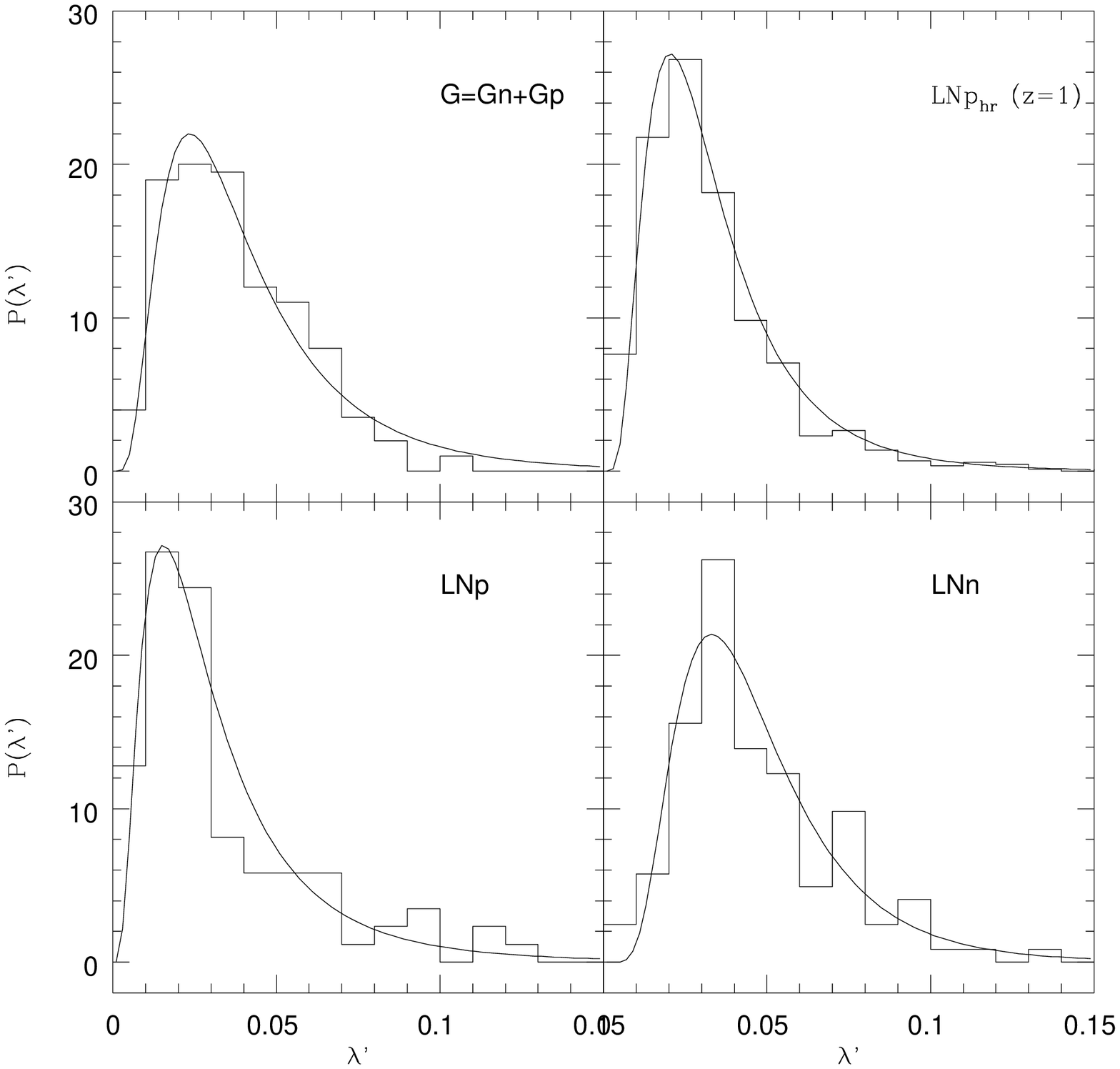}}
\rput[tl]{0}(0.5,2.0){
\begin{minipage}{18.4cm}
  \small\parindent=3.5mm {\sc Fig.}~5.---
Probability distribution (histogram) of the halo spin 
parameter $\lambda'$ for different statistics:
Gaussian (top-left panel), skew-positive and negative lognormal
from the low-resolution simulations (bottom panels),
and skew-positive lognormal from a high-resolution simulation
stopped at $z=1$ (top-right panel). Curves on each panel are lognormal 
best-fits to the data. The $\lambda'_0$ parameter (see eq. [8]) in each 
curve is: 0.035 for the Gaussian (Gp and Gn added), 0.026 for \lnp\ 
(0.029 for the high resolution $z=1$ simulation), and 0.042 for \lnn\
(the $\sigma_{\lambda}$ parameter oscillates around 0.6, see text). 
As can be seen from the plot and confirmed by the $\lambda'_0$
values obtained from the fits, we find that \lnn\ (\lnp) halos
have on average a higher (lower) $\lambda'$ value than their
Gaussian counterparts.
\end{minipage}}
\endpspicture
\end{figure*}

For the halos re-simulated with high resolution, we compute also the 
halo angular momentum distribution (AMD). Figure 6 shows a comparison 
between the AMD of the galaxy-sized halos simulated with the skew-positive 
statistics,  \lnp\ (left panels) and skew-negative, \lnn\ (right panels), 
and the AMD of each one of their Gaussian counterparts. The AMD is 
computed as follows. We find first, for each particle, the specific angular
momentum component along $\mbox{\boldmath $J_{v}$}$, calling it $j$.
We then divide the sphere of radius \rv\ in spherical shells and
each of these shells is in turn divided in four sections. Differences
in the number of particles between cells for most of them are below
a factor of two. The total $j$ of any cell can be
negative because $j$ is a projected component. For each halo,
we measured the mass fraction that there is in cells with negative $j$ and
found that it ranges from 5\% to 50\%. It is remarkable to find a
halo (halo \gtwo) that, with the cell geometry adopted, happens to
have about half of its cells pointing to the opposite direction
of its total angular momentum. The calculation of the AMD
proceeds only for those halos for which this fraction is small, 
less than $\sim $10\%. Cells are ranked according to their $j$ value 
(cells with negative $j$ are rejected) and $M(<j)$ profiles are
then built by counting the cumulative
mass in cells with angular momentum smaller than $j$. Mass
and angular momentum are given in \mv\ and $j_{\rm max}$ units,
respectively, where $j_{\rm max}$ is the maximum value reached by $j$. We
have also plotted in Figure 6 the analytical form of
$M(< j)$ proposed by \cite{Bullock2001} with $\mu = 1.25$; the line is
intended to guide the eye, it is not a fit to our numerical results.
\begin{figure*}[ht]
\pspicture(0.5,-1.5)(15.0,15.5)
\rput[tl]{0}(1.5,15.0){\epsfxsize=16cm
\epsffile{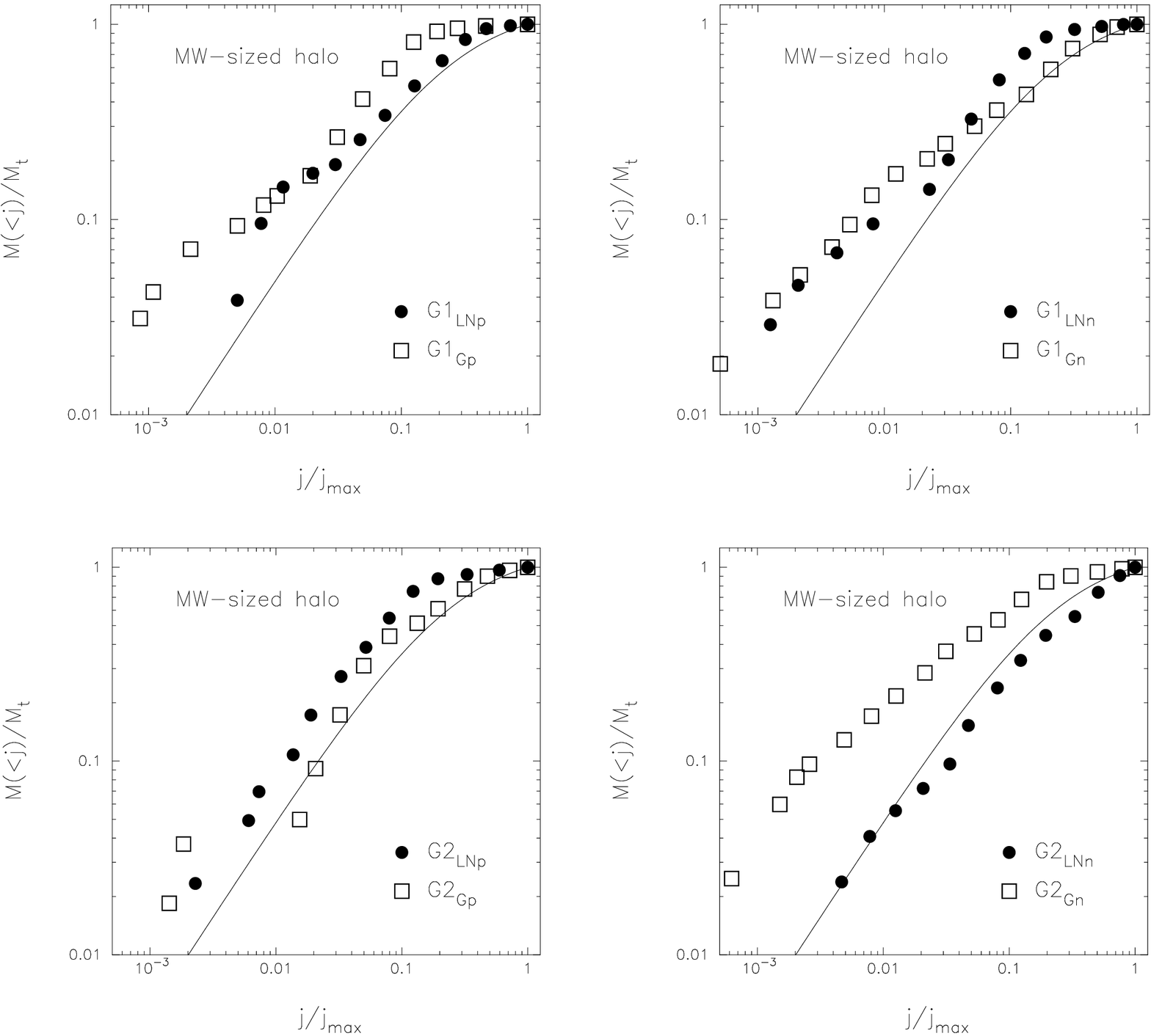}}
\rput[tl]{0}(0.5,0.0){
\begin{minipage}{18.4cm}
  \small\parindent=3.5mm {\sc Fig.}~6.---
Specific angular momentum $j$ distribution of halos
G1 y G2 for the two lognormal statistics \lnp\ and \lnn, and their 
corresponding Gaussian realizations. Only cells with positive $j$ 
were considered and the total
mass $M_t$ was redefined to be the mass contained only in those cells with
positive $j$. The line is taken from the analytical fit of \cite{Bullock2001}
$m(<j) = \mu \tilde{j} / 1 + \tilde{j}$, where $m(<j) \equiv M(<j)/M_{vir}$
and $\tilde{j} \equiv j/j_{max}$, with $\mu = 1.25$. No systematic differences
are seen between halos simulated with lognormal NGICs and those simulated
with the Gaussian statistics.
\end{minipage}}
\endpspicture
\end{figure*}

\subsection{Ellipticities}

We have measured the ellipticities of halos using the tensor of
inertia. This is defined as
\begin{equation}
I_{i,j} = \sum x_i x_j / r^2,
\end{equation}
where the sum is over all particles within $r_{vir}$, $x_i$ ($i=1,2,3$)
are the particle coordinates with respect to the halo center of mass,
and $r$ is the distance of the particle to the halo center.
The ellipticities are then given by
\begin{equation}
e_1 = 1 - \frac{\lambda_1}{\lambda_3},\ \ \ \ \ \ \ \ \ \ \  \ e_2 =
1 - \frac{\lambda_2}{\lambda_3},
\end{equation}
where $\lambda_3 > \lambda_2 > \lambda_1$ are the eigenvalues of the tensor
of inertia. We evaluate the triaxiality parameter using the following
formula (\eg~ \cite{FIdZ90})
\begin{equation}
T = \frac{\lambda_3^2 - \lambda_2^2}{\lambda_3^2 - \lambda_1^2}.
\end{equation}
A halo is prolate (oblate) if $T = 1.0$ ($T = 0.0$).

The ellipticities $e_1$ (solid lines) and $e_2$ (dashed lines)
as a function of radius are
shown in Figure 7 for halos \gfour\ and \gsix\ (left panel)
and for halos \clone\ and \cltwo\ (right panel). For these
halos, no systematic difference is seen in ellipticities
when using different statistics. 
\begin{figure*}[ht]
\pspicture(0.5,-1.0)(15.0,8.5)
\rput[tl]{0}(0.7,8.0){\epsfxsize=18cm
\epsffile{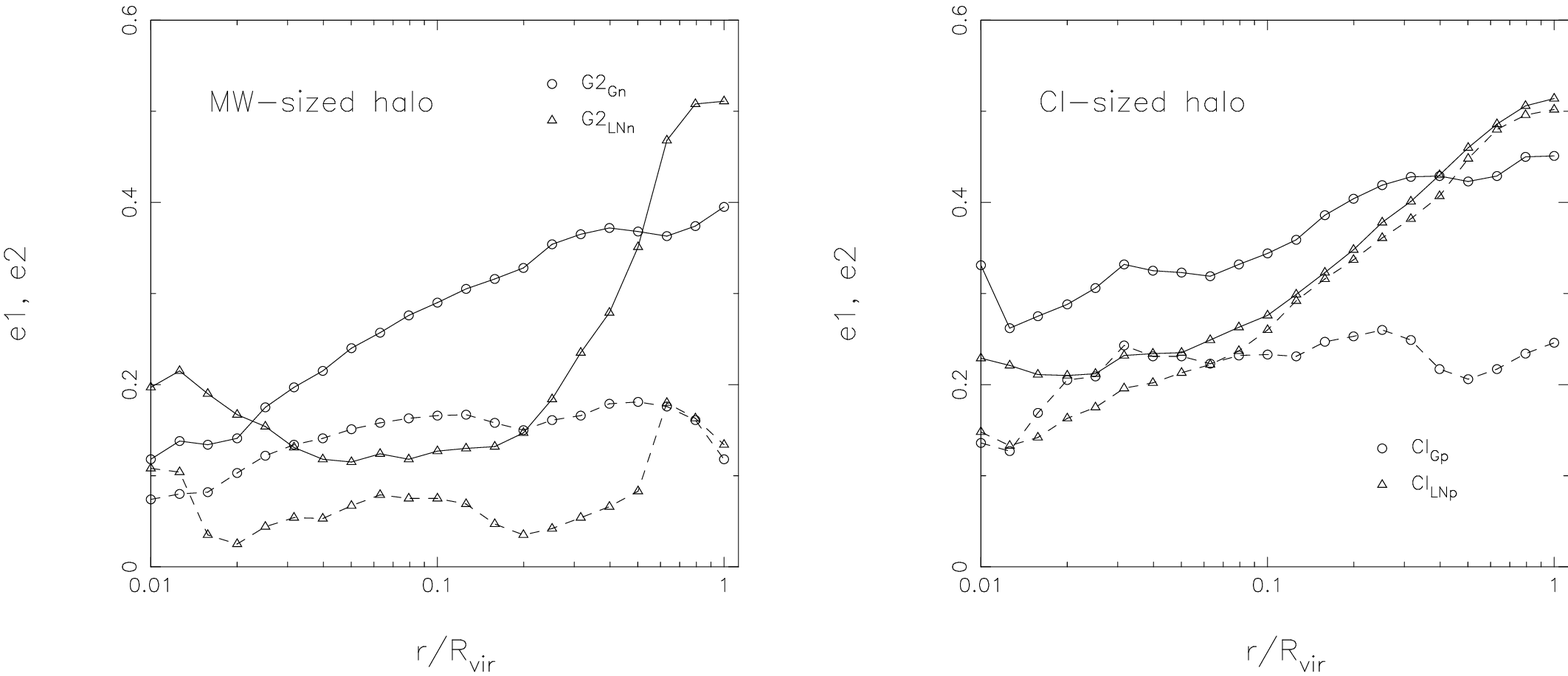}}
\rput[tl]{0}(0.5,-0.2){
\begin{minipage}{18.4cm}
  \small\parindent=3.5mm {\sc Fig.}~7.---
Ellipticities $e1$ and $e2$ as a function of radius 
are shown in left (halo $G2$) and right (halo $Cl$) panels. By definition $e1 <
e2$ and thus $e1$ curves (dashed lines) lie below the corresponding $e2$ 
curves (solid lines).
\end{minipage}}
\endpspicture
\end{figure*}

\subsection{Subhalo Population}

In Figure 8 we show the present-day subhalo \vmax\ cumulative function for
all our galaxy-sized halos displayed in Table 1. All subhalos
inside a radius of  200 \kpch\ from the center of their parent
halo are counted. We plot this function down to $\vmax \sim 20\ \kms$,
where we believe we are complete (\cite{KKVP99}). The observational data 
(five pointed stars) are taken from Klypin et al. Two interesting and
somewhat expected results come just from looking at the figure: substructure 
is suppressed in halos simulated
with a \lnp\ statistics, as compared with the Gaussian one, while
a slightly higher number of subhalos survive in the \lnn\ simulations. These
results can be explained as an effect of the stronger (lighter)
tidal field developed by their parent halos in the corresponding
\lnp\ (\lnn) statistics (see \S 3.1 and \cite{CAV00}).

\section{Summarizing Discussion}

We have analyzed the structure and substructure of \LCDM\ halos obtained
in low and high-resolution cosmological \nbody\ simulations using non-Gaussian 
initial conditions (NGICs). The NGICs were generated in the gravitational 
potential with a lognormal statistics, positive-skewed (\lnp) in one case 
and negative-skewed (\lnn) in the other. 
The sign of the skewness remains the same when passing to the density PDF but 
the statistics are no longer lognormal. Since our aim is to explore in 
a generic way the influence of the NGICs on the properties of the halos,
the particular choice of the NGICs is not relevant. In order to attain a 
fair comparison
with results from the usual Gaussian-initial-conditions simulations, for 
each non-Gaussian simulation, we simulated the corresponding Gaussian one, 
using the same random seed. After the Zel'dovich approximation ($z=50$), the 
skewnesses of the Gaussian, \lnp, and \lnn\ PDFs measured in the simulations
are 0.52, 3.90, and -1.27, respectively, while the $G_3$ parameter values are
3.05 and 0.067 for the \lnp\ and \lnn\ cases, respectively (for the Gaussian 
case  $G_3=1$  by definition, see eq. [1]). These values are after some 
gravitational evolution, and they fall within the range of several 
of the theoretically proposed non-Gaussian statistics.

As discussed in the Introduction, the question of Gaussianity or not in 
the primordial fluctuation field is highly debated in theory as well as
from the point of view of the observations.  Most of the evidence shows that
if the primordial fluctuation field is not-Gaussian, then the deviations from
Gaussianity should not be very large. On the other hand, there is no reason
to {\it a priori} conclude that the statistics is the same at all the scales.
Most of the tests on Gaussianity were applied for relatively large scales (in 
the CMBR anisotropy maps and for cluster abundances). Even in this range 
of large scales, there have been some observational hints that the 
statistics could be different at different scales (e.g., \cite{CNVW03}).
As far as we know, the present paper introduces for the first time a potential
way to test Gaussianity at the smallest scales, using for this a comparison
of the structure of galaxy and cluster-simulated halos with observational
inferences. 

Our main conclusion is that the structure and substructure of the galaxy- and 
cluster-sized \LCDM\ halos are affected by the initial statistics. For the 
skew-positive (skew-negative) statistics, halos are more (less) concentrated,
cuspier (shallower), and with less (more) substructure than the corresponding
Gaussian halo counterparts. These main findings are clearly seen in Figs. 2 and 
3. 

According to our results, the skew-positive NGICs would exacerbate
the difficulty that the Gaussian \LCDM\ model faces with observations: the 
rotation curves of dark-matter dominated (dwarf and low surface
brightness) galaxies show that the inner density profiles of the halos
are shallower and less concentrated than those of the \LCDM\ halos 
(see \cite{CAVF02} for references). We notice that tests related to 
cluster abundances and their evolution suggested skew-positive deviations 
from Gaussianity (see Introduction).
On the other hand, the skew-negative NGICs work in the direction 
of making the galaxy-sized dark halos less concentrated and shallower 
than the Gaussian ones. 
\begin{figure*}[ht]
\pspicture(0.5,-1.5)(15.0,20.0)
\rput[tl]{0}(2.0,19.5){\epsfxsize=15.5cm
\epsffile{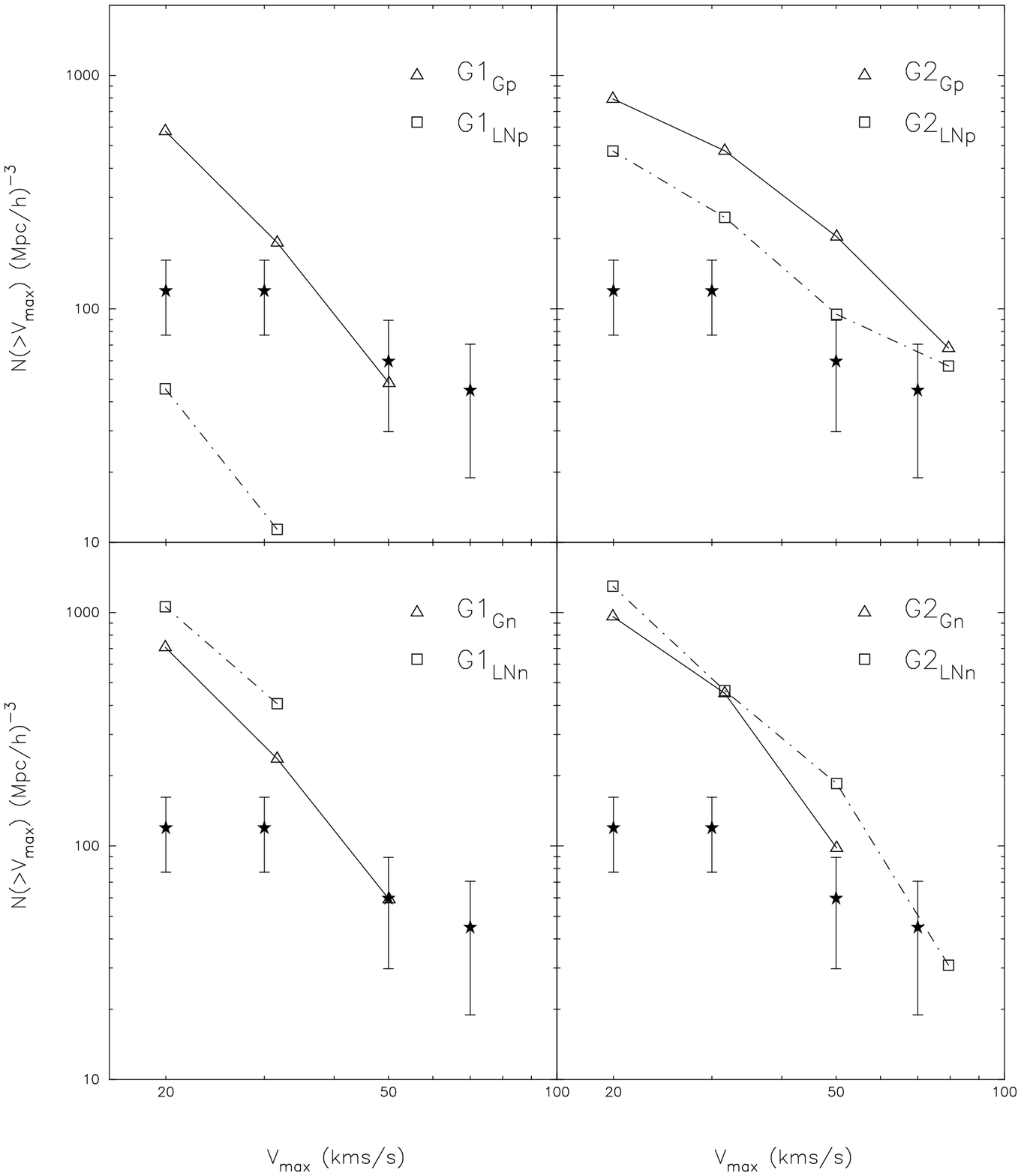}}
\rput[tl]{0}(0.5,1.0){
\begin{minipage}{18.4cm}
  \small\parindent=3.5mm {\sc Fig.}~8.---
Cumulative subhalo $\vmax-$functions of halos G1 y G2 for
the Gaussian and lognormal statistics. All subhalos with \vmax\ greater 
than 20 \kms\ within a sphere of radius 200 \kpch\ were counted. The averaged
$\vmax-$function from satellites of the Milky Way and Andromeda is represented by stars 
(taken from \cite{KKVP99}). We plot for clarity only the error bars for the 
``observed'' circular velocity function. It is clear that substructure 
is erased more efficiently in halos with a \lnp\ 
statistics than with a Gaussian one (see top panels). 
This effect can be traced back to the stronger tidal field experimented by 
subhalos in the \lnp\ scenario. The opposite effect is expected when the 
amount of substructure in a \lnn\ scenario is compared with its Gaussian 
counterpart. This is indeed the case: more subhalos survive under the \lnn\ 
statistics than under the Gaussian one.
\end{minipage}}
\endpspicture
\end{figure*}

Regarding the subhalo population, since the 
halos in the \lnp\ statistics are very concentrated, more subhalos are tidally
destroyed than in the Gaussian case. The opposite occurs with the 
subhalo population in the \lnn\ simulations, although the difference with 
the Gaussian results in this case is very small. Observations show
that the Gaussian \LCDM\ model overpredicts the number of satellite subhalos 
within simulated galaxy-sized CDM halos (\cite{KKVP99}; \cite{Moore99a}). 
Nevertheless, this difference could be due to astrophysical processes 
(reionization, for example) which inhibit the formation of dwarf galaxies 
within the subhalos (\cite{BKW2000}, \cite{Benson}). 

The halo spin parameter probability 
distribution also has shown to depend on the primordial statistics:
\lnn\ (\lnp) halos have on average a higher (lower)
$\lambda'$ value than their Gaussian counterparts. A potential difficulty
pointed out for the Gaussian halos is that the disks formed within 
them result too much concentrated (\cite{NW94}). A possible
way to improve on this problem is to generate halos with larger
spin parameters than the Gaussian CDM ones. For example, this is the 
case for the \lnn\ model presented here. 

From a theoretical point of view, our results raise an interesting problem:
the  density profile of the dark matter halos seems to depend on initial 
conditions. The question whether the structure of CDM halos is universal 
and non-dependent on cosmology, power spectrum, formation history, etc.
has been several times discussed in the literature. 
Avila-Reese et al. (1998, 1999) have shown that the CDM halo density profiles
oscillate around the NFW profile, depending mainly on the halo mass aggregation
history (see also \cite{WBPKD02}). Nevertheless, a significant variation
of the {\it inner} density profile was not expected, unless thermal processes
are included, as is the case of self-interacting dark matter (e.g., \cite{CAVF02})
or high rms-velocity dispersions due to thermal relicts (e.g.,
Avila-Reese et al. 1998, 2001; {\L}okas \& Hoffmann 2000). It has been also 
shown that variations in the power spectrum shape
and the type of the halo collapse (hierarchical merging or by fragmentation) do not
introduce significant changes in the halo density profiles (\cite{Moore99b};
\cite{HJS99}; \cite{ACV01}; \cite{Man03}). The fact that the
initial statistics of the fluctuation field does significantly affect the
inner structure of the CDM halos is somewhat unexpected although, using 
the semi-numerical method of Avila-Reese et al. (1998), it was already shown
that for a possitive lognormal statistics (in the density field), the halos
are more cuspy than in the Gaussian case (\cite{Viniegra97}; \cite{Avila98}).
Halos form typically from the high-density peaks in the matter distribution.
In the skew-possitive statisitics, high-density peaks are 
more abundant, isolated and ``coherent'' than in the Gaussian statistics;
the opposite happens with the  skew-negative statistics.
The final structure of the halos is probably related to these facts. 
A deeper analysis of this question has been deferred to a future paper. 

In the last few years, the observational determination of the cosmological 
parameters attained an unprecedent high accuracy and the body of astronomical
observations of the local and high-redshift universe has grown dramatically.
This progress not only allows to constrain better the cosmological models, but 
also opens a new dimension in the relaxation of some hypotheses made before,
as well as in the exploration of second-order phenomena in the cosmic
structure formation process. An example of this is just the statistics of the 
initial fluctuation field. Having now well defined the main cosmological 
parameters and a succesful model of cosmic structure formation, one may explore
the influence of NGICs on structure formation free of degeneracies
with other parameters and assumptions (e.g., with $\Omega_m$ and $\sigma_8$).

\acknowledgments
 
We are grateful to Anatoly  Klypin and Andrey Kravtsov for kindly 
providing us a copy of the ART code in its version of multiple mass, and for
our enlightening discussions. PC is grateful to A. Kravtsov for providing access
to a computer where the $256^3$-particle simulation
was run. GP is grateful to Instituto de Astronom\'{\i}a, 
UNAM, for its kind hospitality during the development of this work. Part of the 
ART simulations were performed at the Direcci\'on General de Servicios de
 C\'omputo Acad\'emico, UNAM, using an Origin-2000 computer. This work was 
supported by CONACyT grants 33776-E to V.A. and 36584-E to P.C.



\end{document}